\documentclass[a4paper,11pt]{article}
\usepackage{pos}

\title{Theory overview on neutrino studies and outlook}

\author*[a]{Toni M{\"a}kel{\"a}}

\affiliation[a]{Department of Physics and Astronomy, University of California, Irvine, CA 92697 USA}

\emailAdd{tmakela@uci.edu}

\abstract{The forward neutrino program at the Large Hadron Collider has entered the era of providing the first measurements and observations. As it has notable connections to astrophysics and will complement the results and projections of key measurements at the LHC and other measurements, it is becoming increasingly important for collider physicists to recognize and utilize the unique potential of this novel research direction. The present work reviews a selection of questions related to the nature, masses and interactions of neutrinos, and highlights some of the recent projections for the LHC Run 3 as well as the high-luminosity run.}

\FullConference{The Thirteenth Annual Large Hadron Collider Physics (LHCP2025)\\
5-9 May 2025\\
Taipei\\}

\begin{document}
\maketitle

Neutrinos are among the most interesting topics of contemporary particle phenomenology. Although they are established standard model (SM) particles, a full understanding of them requires physics beyond the SM (BSM). Foremost, there are no right-handed neutrinos, and consequently no neutrino mass term, in the SM, but masses are required to explain solar and atmospheric neutrino oscillations. Several questions in neutrino physics hence surround the nature and origin of their masses: what is the absolute mass scale, and is the ordering normal or inverted? Are there right-handed neutrinos, or are the neutrinos Majorana instead of Dirac fermions, or is there a Majorana mass component? This would lead to breaking $SU(2)_L$ invariance and lepton number symmetry. Besides understanding neutrino masses, it is important to characterize their interactions and oscillations precisely. For instance, neutrinos might explain the baryon asymmetry via leptogenesis, and leptonic CP violation is probed in three-flavor oscillation studies. Albeit too light and fast to be dark matter candidates, SM neutrinos are an important background in dark matter searches, and have important connections to dark matter in many BSM theories via their interactions and possible sterile neutrinos. It is also interesting to ask if quark and lepton mixing are related by some flavor symmetry, and some models remain to be tested with oscillation data~\cite{Chauhan:2023faf}. Despite the inexhaustiveness of this list, neutrinos are clearly of great relevance to cosmology and astrophysics (cf. Refs.~\cite{Lesgourgues:2006nd,Hannestad:2010kz,Abazajian:2016hbv,Escudero:2024uea}). While there is much to be done for the many on-going and planned experiments across the globe, the focus here is on the novel neutrino program at the Large Hadron Collider (LHC). A brief review of select theory topics is followed by a summary of some challenges and prospects of LHC neutrino studies.

Having no electric charge, neutrinos are the only SM fermions that could be their own antiparticles. In this case, the charge conjugation operator merely turns left-handed chirality into right-handed, $\nu_L=(\nu_L)^C=\nu^C_R$, or vice-versa. This introduces a Majorana mass term of the form 
$\frac{-m}{2}
\left(   \overline{\nu_L} \mathbf{C}^{-1} \overline{\nu_L}^T
         + \nu_L^T \mathbf{C}^{-1} \nu_L^T
\right)$, which can be constructed using only left or right handed states, and violates lepton number. The question of whether a $\nu$ or a $\overline{\nu}$ was produced and observed in an experiment is then replaced by the problem of which handedness it had at the production and detection sites, determining how it appears to interact with SM particles. Although a suppression of the form $(m/E)^2$ hinders direct observations of chiral oscillations at relativistic energies, Majorana effects can be significant at low energies~\cite{Li:2023iys}. At high energy colliders, Majorana effects might however be probed indirectly by searching for lepton number violating processes~\cite{Deppisch:2015qwa}. The main test for Majorana nature is nonetheless the low-energy neutrinoless double beta decay, seen e.g. as the proton number of a decaying nucleus $N$ increasing by two, 
$~^A_Z N \to ~^{~~A}_{Z+2}N + 2e^-$~(cf. Ref.~\cite{Rodejohann:2011mu}). 

Alternatively, adding a right-handed neutrino $\nu_R$ to the SM allows introducing a Dirac mass term. The $\nu_R$ must be sterile as they cannot have direct weak interactions; otherwise they should contribute to the invisible $Z$ decay width. If stable, such sterile neutrinos are also a dark matter candidate. They are SM gauge singlets, and can have a mass matrix with Majorana and Dirac terms, with $m_N$ as the heavy sterile neutrino mass and $m_D$ connecting the left- and right-handed neutrinos. If $|m_N|\gg|m_D|$, this results in the type I seesaw model with $m_\nu \approx -m_D m_N^{-1} m_D^T$. If $|m_D|\gg|m_N|$ instead, the neutrinos are quasi-Dirac, but lepton number is not exactly preserved. Searches for sterile neutrinos typically probe $m_N$ and look for active-sterile neutrino mixing.

In the SM effective field theory, Majorana mass is introduced 
at dimension 5 by the Weinberg operator~\cite{Weinberg:1979sa} 
$\epsilon_{ij}\epsilon_{kl} H_i H_k L_j C L_l / \Lambda_5$,
where the $\epsilon_{ij}$ are dimensionless couplings, $\Lambda_5$ is the energy scale of new physics, $H_i$ is the Higgs doublet, $L_j$ the left-handed lepton doublets, and $C$ the charge-parity conjugation operator. For example, this can arise from introducing $SU(2)_L$ triplet scalars (type II seesaw) or fermions (type III), in which case there is potential for a discovery at high-energy colliders due to the exotics coupling to the electroweak bosons. In radiative mass models, small Majorana masses are generated at loop level, e.g. by generalizing the Weinberg operator to $LLHH(H^\dagger H)^n,~n=0,1,2,...$, or by using non-Weinberg operators corresponding to diagrams where some legs can be closed into loops, forming neutrino self-energy diagrams. Having SM particles in these loops could then imply non-standard interactions (NSI).

Until recently, neutrino measurements have been performed at either very small or extremely high energies, using neutrinos produced at reactors ($<$MeV) or low-energy accelerators (GeV), solar neutrinos (10 MeV), and astrophysical observations (EeV). The medium-to-high-energy gap is bridged by the recent forward neutrino program at the LHC, currently comprising of the FASER~\cite{FASER:2018ceo,FASER:2018bac,FASER:2019dxq,FASER:2020gpr,FASER:2022hcn} and SND@LHC~\cite{SHiP:2020sos,SNDLHC:2022ihg} collaborations, which have already provided the first neutrino observations~\cite{FASER:2023zcr,SNDLHC:2023pun} and measurements~\cite{FASER:2024hoe,FASER:2024ref}. Both are to resume operations during Run 4, with upgraded detectors~\cite{FASER:2025myb, Abbaneo:2926288}. These experiments, together with novel detectors proposed at the FPF~\cite{ANCHORDOQUI20221,Adhikary:2024nlv,Feng:2022inv}, or at the surface exit points of the LHC neutrino beam~\cite{Ariga:2025jgv,Kamp:2025phs}, probe TeV scale neutrinos, whereas the recently approved SHiP~\cite{Ahdida:2023okr,SHiP:2025ows} will contribute to the CERN neutrino program at more modest energies of order 10~GeV. It is important to ensure that neutrino measurements can be performed at a variety of energies also in the future. For instance, searches for sterile neutrinos, or heavy neutral leptons for generality, are a direct probe of neutrino mass generation and should be performed at all scales, and can be done at FASER, the FPF, and SHiP via meson decays~\cite{Kling:2018wct,Feng:2024zfe,Adhikary:2024nlv,Wang:2024mrc}. To reap maximal benefits from the forward neutrino program, it is also important to further develop the projections for these experiments as well as their detection capabilities. This will require for example integrating state-of-the-art SM and BSM calculations into Monte Carlo (MC) computations, and improving our understanding of the production of forward hadrons giving rise to neutrino fluxes. 

While the neutrinos at SHiP originate from the decays of hadrons produced as the Super Proton Synchrotron beam hits a fixed target, the ones observed at FASER, SND@LHC and the FPF arise from the decays of forward hadrons produced in the pp collisions at the ATLAS interaction point (IP). These neutrinos, and possible long-lived particles, aren't observed at the IP, but in forward detectors removed hundreds of meters along the line of sight from the IP. As the spectra of observed $\nu$ interactions are not only altered by physics effects at detection, but also at the production site, it is essential to understand the uncertainty of the incoming neutrino flux. This arises dominantly from our ignorance of which models for forward hadron production at the IP give the best description of nature; hence a description of the flux requires comprehending forward hadron composition. The distributions of light mesons ($m < 1$~GeV) are simulated using various event generators developed for cosmic ray and forward LHC physics~\cite{Ahn:2009wx,Ahn:2011wt,Riehn:2015oba,Fedynitch:2018cbl,Pierog:2013ria,Roesler:2000he,Ostapchenko:2010vb,Fieg:2023kld,Fedynitch:2015phd}. 
In contrast, the charm mass ($>1$~GeV) gives a marginal hard scale, and perturbative predictions (up to NLO) exist for charmed hadron production using either collinear or $k_T$ factorization~\cite{Ahn:2009wx,Ahn:2011wt,Riehn:2015oba,Fedynitch:2018cbl,Buonocore:2023kna,Bai:2020ukz,Bai:2021ira,Bai:2022xad,Bhattacharya:2023zei,Maciula:2022lzk}. 
These predictions yield MC samples of $\nu$ flavor, position, and 4-momenta, which are used in Ref.~\cite{Kling:2023tgr} to assess the ultimate theory uncertainty in the neutrino flux via a Fisher information approach accounting for correlations between the predictions. As experimental data becomes available, the forward neutrino observations will test which models provide the best descriptions in the forward region, and help tune MC generators.

\begin{figure}[htb]
\centerline{%
\includegraphics[width=0.32\textwidth,trim={0mm 0mm 35mm 0mm},clip]{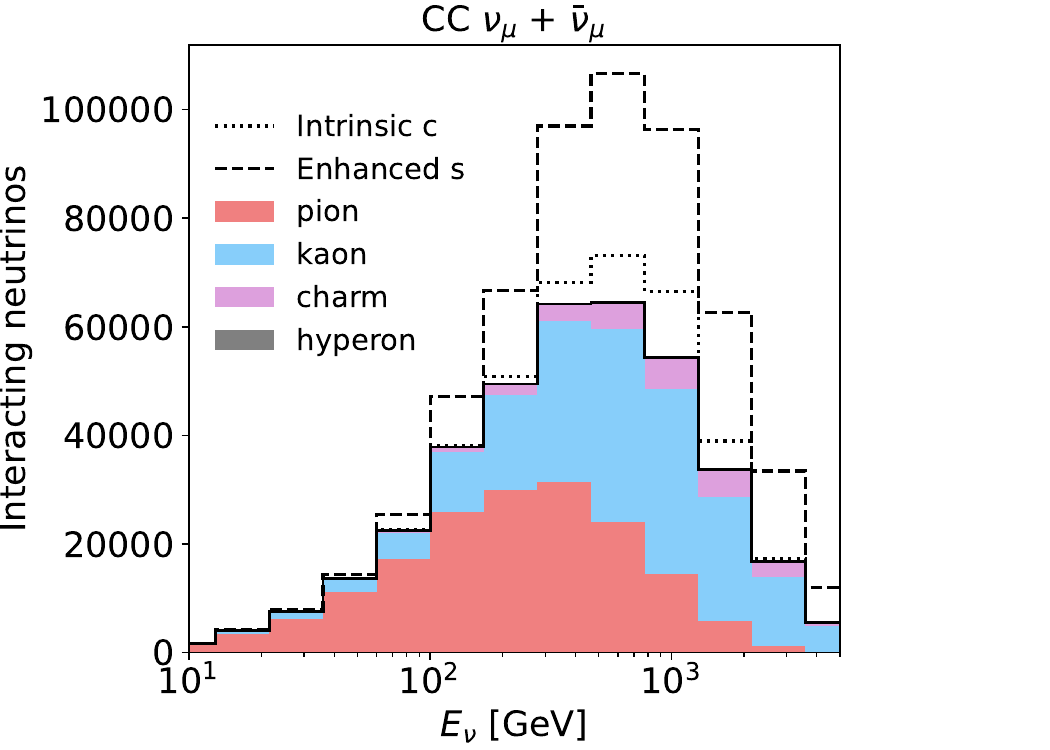}
\includegraphics[width=0.32\textwidth,trim={0mm 0mm 35mm 0mm},clip]{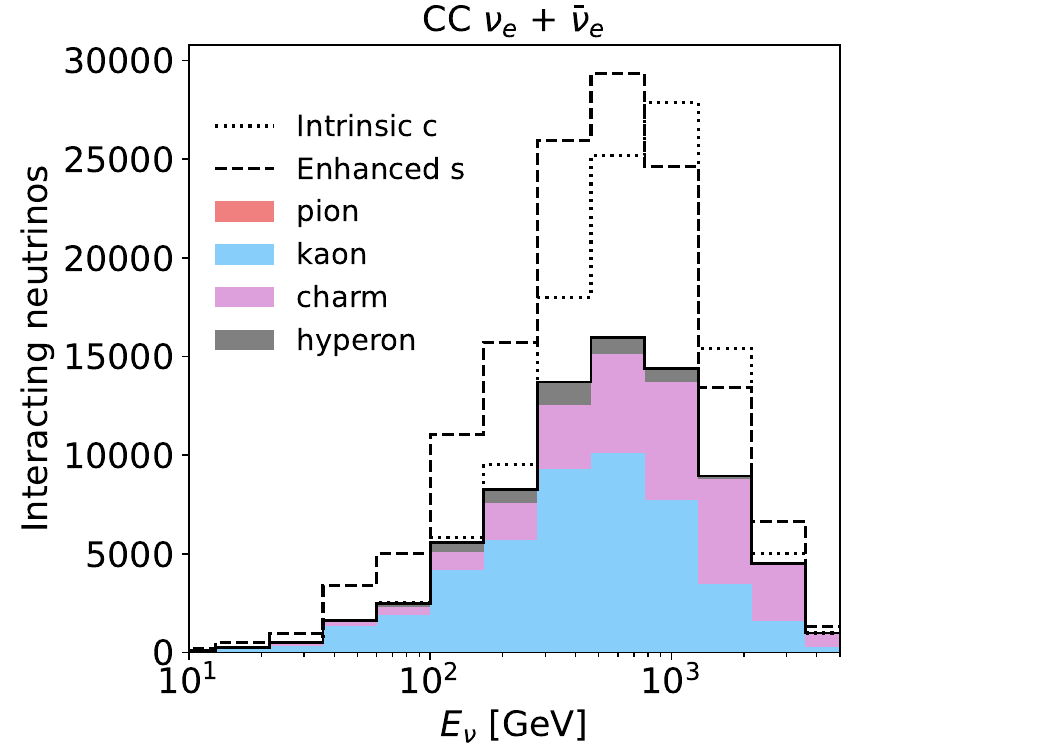}
\includegraphics[width=0.32\textwidth,trim={0mm 0mm 35mm 0mm},clip]{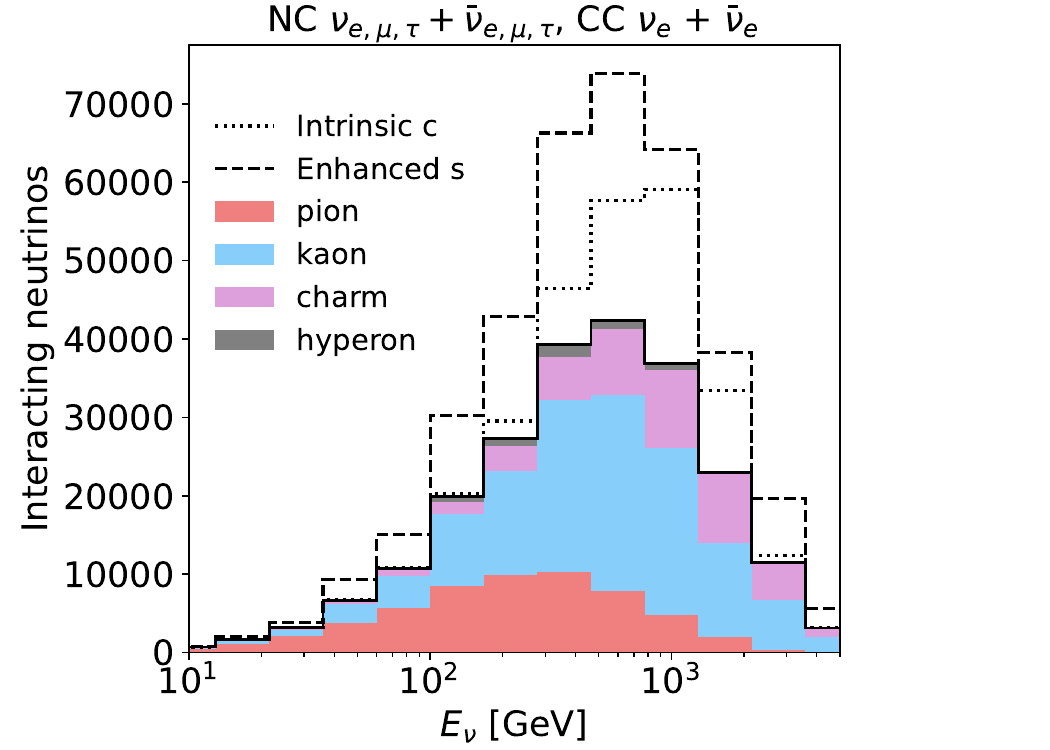}}
\caption{The composition of CC interaction spectra for incoming $\nu_\mu+\overline{\nu}_\mu$ (left) and $\nu_e+\overline{\nu}_e$ (middle) for a hypothetical detector assuming perfect lepton identification. 
This constrains forward hadron production, as pions do not contribute to the $\nu_e$ spectrum consisting mainly of kaons and charmed hadrons. 
Further constraining power can be obtained with reliable final state lepton charge identification, as hyperons decay only to $\overline{\nu}_e$ ~\cite{Kling:2025lnt}. Additionally, the characteristic changes induced on the spectra by enhanced strangeness production and proton intrinsic charm models are illustrated qualitatively. The consequences of poor lepton identification are depicted for a case where low-energy $\nu_e$CC events are indistinguishable from neutral current (NC) interactions, and $\pi$ contributions leak into observed $\nu_e$ spectra via $\nu_\mu$NC (right). Image after Ref.~\cite{Ariga:2025jgv}.}
\label{Fig:composition}
\end{figure}

Ideally, the experiments measure $\nu$CC interactions for all neutrino flavors in bins of energy and radial distance from the line-of-sight (LOS). All $\nu_\tau$, and some high-energy $\nu_e$, arise from charmed hadron decays, which are typically more spread out along the LOS than light hadrons. The remainder of the $\nu_e$ arise mainly from kaons, which also contribute to the $\nu_\mu$ spectrum. The rest of the $\nu_\mu$ originate chiefly from pion decays. Sensitivity to the energy and rapidity distributions of all $\nu$ flavors then enables constraining forward hadron production. Further, future experiments should aim for the best possible lepton charge identification capabilities, allowing the distinction of $\nu$ and $\overline{\nu}$ events~\cite{Kling:2025lnt}. Precise lepton charge identification may also yield the first direct detection of $\overline{\nu}_\tau$ (separate from $\nu_\tau$), the last unobserved SM particle, at the high-luminosity LHC~\cite{Adhikary:2024nlv}. Moreover, effects like proton intrinsic charm, or enhanced strangeness production, can be identified by their characteristic alterations to the expected neutrino spectra~\cite{Adhikary:2024nlv}. However, if detector capabilities are insufficient, the constraining power is reduced~\cite{Ariga:2025jgv}. This is summarized in Fig.~\ref{Fig:composition}.

The LHC neutrino program is also relevant for astrophysics, and may help solve long-standing issues such as the cosmic ray (CR) muon puzzle: a deficit of high-$E$ muons in air shower simulations (QCD) vs measurements, first observed at the Pierre Auger Observatory~\cite{PierreAuger:2014ucz,PierreAuger:2016nfk,PierreAuger:2021qsd}. 
Presently, this has a combined statistical significance of $8\sigma$, cf. Ref.~\cite{Albrecht:2021cxw} for review.
The high energies of the incident cosmic rays, starting at $E \sim 10^8$~GeV, translate to $\sqrt{s} \simeq \sqrt{2Em_p} \simeq 14$~TeV in the pp collision center-of-mass frame. 
Therefore, the high-energy forward hadrons produced at the LHC, and probed by the forward neutrino experiments, test QCD in a kinematic regime corresponding to secondary particle distributions in CR showers. According to the \textit{enhanced strangeness hypothesis}, an increase in strange hadron production yields higher numbers of muons, leading to less pions and more kaons. By reweighting the counts of neutrinos associated with pions by $(1 - f_s)$, and those coming from kaons by $(1 + F f_s)$, with $F$ a phenomenological factor accounting for the difference in $\pi / K$ production rates, Ref.~\cite{Anchordoqui:2022fpn} found that a value of $f_s=0.5$ could explain the CR muon issue. Such values can already be probed by FASER$\nu$~\cite{Kling:2023tgr} during Run 3. Scenarios where $f_s$ is not constant, but has lower values at LHC energies, will be constrained by the FPF~\cite{Kling:2023tgr}.

\begin{figure}[htb]
\includegraphics[width=0.22\textwidth]{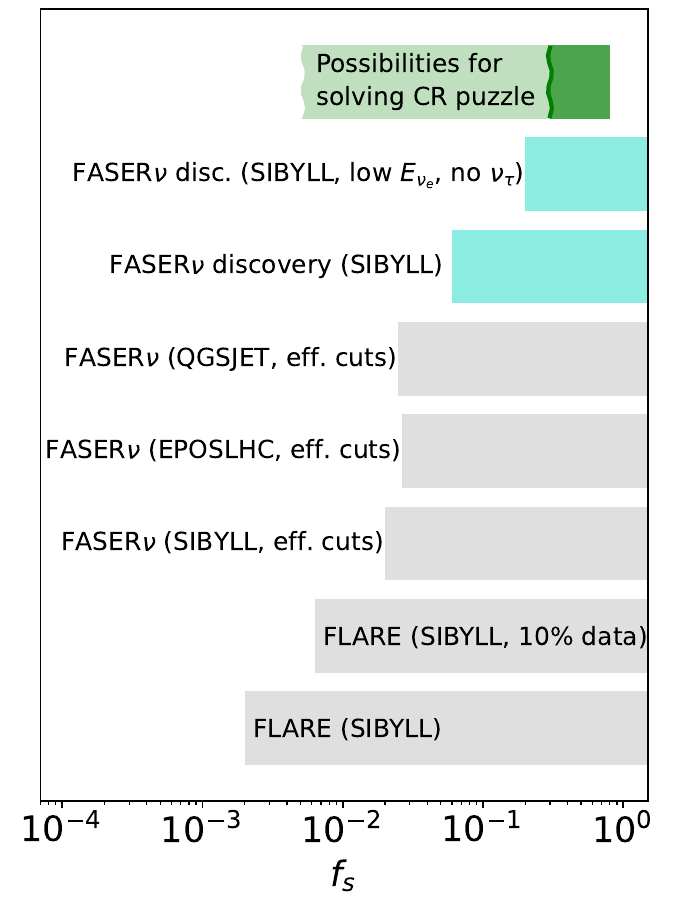}
\includegraphics[width=0.44\textwidth]{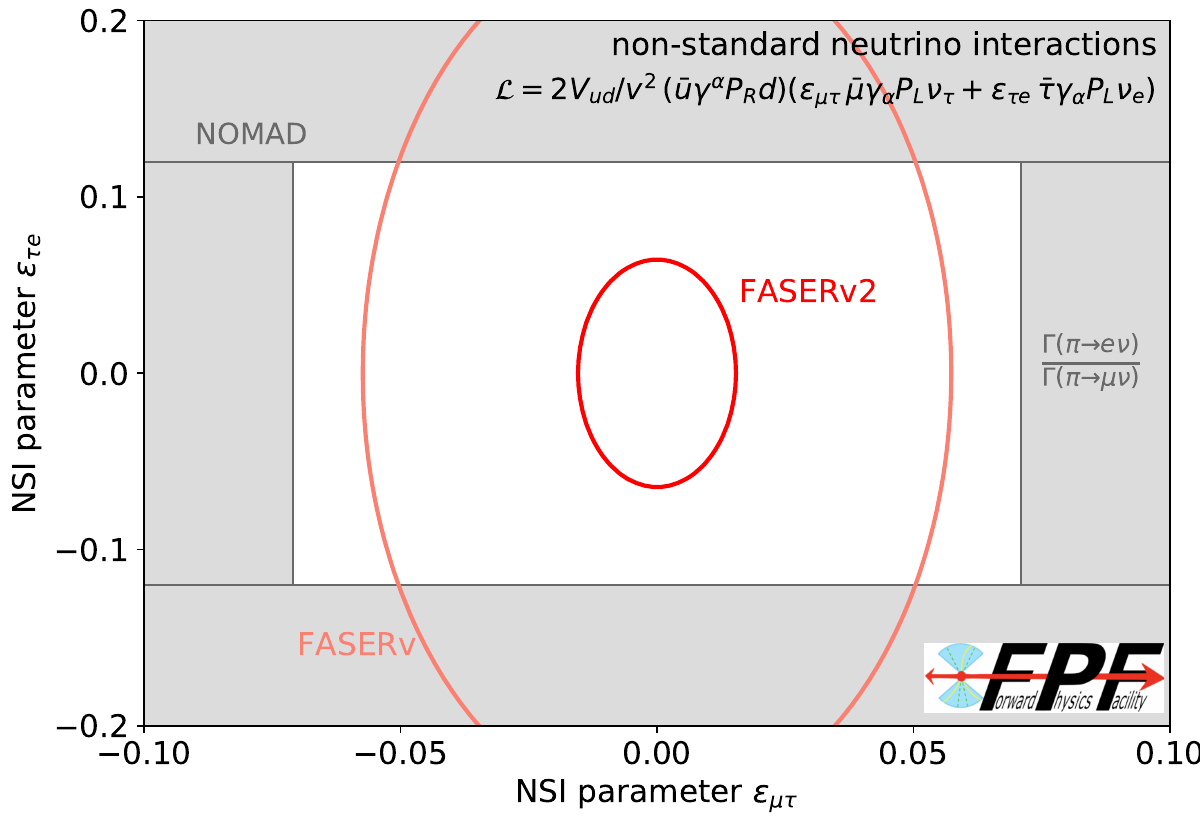}
\includegraphics[width=0.32\textwidth]{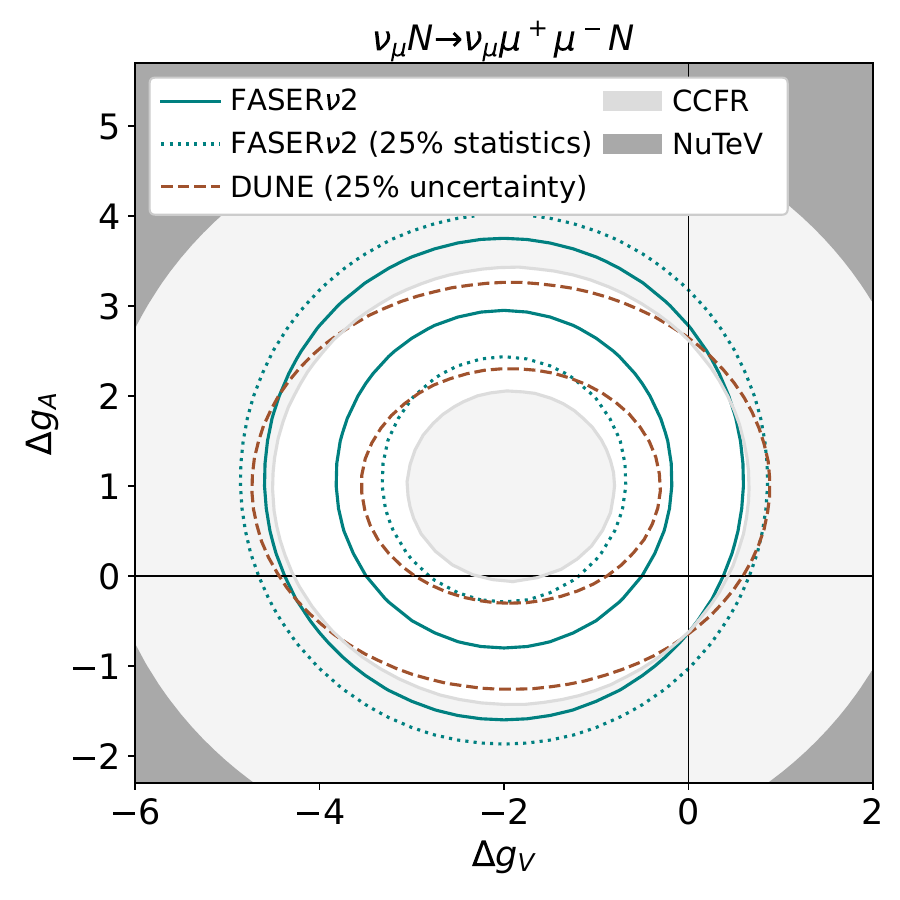}
\caption{
\textit{Left:} 
$2\sigma$ constraints (gray) for $f_s$ at FASER$\nu$ and FPF, and the FASER$\nu$ discovery potential (turquoise), with and without $\nu_\tau$ and high-$E$ $\nu_e$ contributions. All bounds cover the region favored by the enhanced strangeness solution to the muon puzzle (dark green), though the effect may be more subtle at the LHC (light green)~\cite{Kling:2023tgr}.
\textit{Middle:} Projected FASER$\nu$(2) sensitivities to NSI operators violating lepton flavor universality~\cite{Kling:2023tgr, Adhikary:2024nlv}. 
\textit{Right:} 95\% confidence level sensitivities to axial and vector coupling modifications due to NSI affecting trident measurements. The DUNE bounds (dashed brown) assume a 25\% uncertainty cross section measurement. The FASER$\nu$2 bounds (teal) are based on 100\% and 25\% of expected data. The NuTeV bound is found assuming a BSM/SM cross section ratio equal to CCFR due to similar energy~\cite{Altmannshofer:2024hqd}.}
\label{Fig:BSM}
\end{figure}

The neutrino spectra observed in forward neutrino experiments can be affected by BSM effects either at the \textit{production} or \textit{detection} sites. Such phenomena can be studied in the context of NSI. Consider for instance adding dimension-6 operators~\cite{Falkowski:2021bkq}, so that
\begin{equation*}
\mathcal{L} = \mathcal{L}_{\rm SM}
- \frac{2 V_{ud}}{v^2}
  (\bar{u}\gamma^\kappa P_Rd)
  \left[  \epsilon_R^{\mu\tau} (\bar{\ell}_\mu \gamma_\kappa P_L \nu_\tau)
        + \epsilon_R^{\tau e}  (\bar{\ell}_\tau\gamma_\kappa P_L \nu_e   )
  \right],
\end{equation*}
where $\epsilon_R^{\mu\tau}$ changes the incoming $\nu_\tau$ flux by introducing a new pion decay channel, while $\epsilon_R^{\tau e}$ modifies the rate of observed $\tau$ leptons in the detector. The ultimate constraints achievable for these operators at FASER$\nu$ and the FPF are shown in Fig.~\ref{Fig:BSM} middle. Furthermore, FASER$\nu$2 has the potential for a first conclusive 5$\sigma$ observation of neutrino tridents, a $\nu_\mu N \to \nu_\mu \mu^+ \mu^- N$ process, proven to be notoriously difficult experimentally~\cite{Altmannshofer:2024hqd}. While measuring the SM process is a triumph in itself, it can also be used for investigating BSM interfering with muon interactions by modifying 4-Fermi vector and axial couplings~\cite{Altmannshofer:2024hqd, Altmannshofer:2019zhy}. Here the high energies of the LHC neutrinos allow complementing DUNE bounds despite limited statistics, as shown in Fig.~\ref{Fig:BSM} right~\cite{Altmannshofer:2024hqd}.

Lastly, forward neutrinos can constrain global parton distribution functions (PDF), as discussed in Ref.~\cite{Cruz-Martinez:2023sdv} in the context of the FPF at the LHC, and in Ref.~\cite{MammenAbraham:2024gun} for the Future Circular Collider. At the latter, pPb collisions may probe nuclear PDFs even down to $x\sim 10^{-9}$. At the FPF, most improvement is expected for the valence and strange quarks: constraining the valence (strange) quark PDFs benefits from good lepton charge identification (charm tagging). The statistics at the Run 3 experiments are, however, deemed insufficient for constraining PDFs, further motivating the FPF and electron-ion collider efforts. Reducing PDF uncertainties is relevant to key measurements at the LHC, such as the inclusive Drell-Yan cross section, the $W$ mass, and the Weinberg angle. In summary, a strong neutrino program at the LHC and future colliders will not only increase our understanding of neutrinos, but will also benefit other experiments, both at colliders and within other fields of fundamental physics.

\end{document}